\documentclass[preprint,prb,amsmath,superscriptaddress,aps,subeqn,showpacs]{revtex4}
\usepackage{graphicx}

\begin{document}

\title{Electron Transport Properties of Bilayer Graphene}

\author{X. Li}
\affiliation{Department of Electrical and Computer Engineering, North Carolina State University,
Raleigh, NC 27695-7911}

\author{K. M. Borysenko} \affiliation{Department of Electrical and Computer Engineering, North
Carolina State University, Raleigh, NC 27695-7911}

\author{M. Buongiorno Nardelli} \affiliation{Department of Physics, North Carolina State University,
Raleigh, NC 27695-8202} \affiliation{CSMD, Oak Ridge National Laboratory, Oak Ridge, TN 37831}

\author{K. W. Kim} \affiliation{Department of Electrical and Computer Engineering, North Carolina
State University, Raleigh, NC 27695-7911} 

\begin{abstract}

Electron transport in bilayer graphene is studied by using a first principles analysis and the Monte Carlo simulation under conditions relevant to potential applications.  While the intrinsic properties are found to be much less desirable in bilayer than in monolayer graphene, with significantly reduced mobilities and saturation velocities, the calculation also reveals the dominant influence of extrinsic factors such as the substrate and impurities.  Accordingly, the difference between two graphene forms are more muted in realistic settings although the velocity-field characteristics remain substantially lower in the bilayer.  When bilayer graphene is subject to an interlayer bias, the resulting changes in the energy dispersion lead to stronger electron scattering at the bottom of the conduction band.  The mobility decreases significantly with the size of the generated bandgap, whereas the saturation velocity remains largely unaffected.


\end{abstract}


\maketitle

\section{Introduction}
Graphene has received much attention in the last few years due to
its unique properties. In addition to the significant interests in
fundamental physics, stemming in part from the relativistic-like
behavior of the massless charge particles around the Dirac cone,
this material is very attractive in many applications, particularly
in high speed devices.~\cite{Novoselov2004,Novoselov2005, Lin2010,
Neto2009} However, the gapless electron spectrum of monolayer
graphene (MLG) makes it difficult to turn off the electrical current due to tunneling.  Bilayer graphene (BLG), on the other hand, can provide a
finite band gap up to hundreds of meV, when the inversion symmetry
between top and bottom layers is broken by an applied perpendicular
electric field.~\cite{McCann2006,Castro2007,Min2007,Zhang2009} A
current on/off ratio of about 100 was observed at room
temperature,~\cite{Xia2010} offering a much needed control for
nonlinear functionality. Unfortunately experimental studies have also
indicated that the typical mobility of charge carriers in BLG
may be substantially smaller than in MLG.~\cite{Xiao2010} A recent
work based on the first principles calculations has suggested that
this discrepancy may start with the intrinsic transport properties,
which results from substantial differences in electron-phonon
coupling in these two materials.~\cite{Borysenko2011} Additionally,
it is widely accepted that extrinsic factors such as charged
impurities, disorder, and surface polar phonons (SPPs) can significantly alter carrier transport in graphene on a substrate, a most commonly used configuration.~\cite{Jang2008,Sarma2010,Fratini2008} Despite extensive research efforts, a comprehensive understanding of electron transport properties in BLG is still a work in progress.

The purpose of the present paper is to address this issue theoretically by taking advantage of a first principles analysis and the Monte Carlo simulation.  Specifically, the impacts of substrate conditions and perpendicular electric fields are examined.  The calculation results indicate that graphene in the bilayer form loses much of its advantage over conventional semiconductors in the low field transport, particularly when the band structure is modified to induce non-zero energy gap.  The saturation drift velocity, on the other hand, can remain relatively high.  Due to the screening, electrons in BLG appear less susceptible to the interactions with remote Coulomb sources, such as SPPs and impurities on the substrates, compared to the monolayer counterparts.  Below, we begin with a brief overview of the models used to estimate the relevant scattering rates.

\section{Relevant Scattering Mechanisms}

\begin{center}
\begin{figure}
\includegraphics[bb=243 514 416 672]{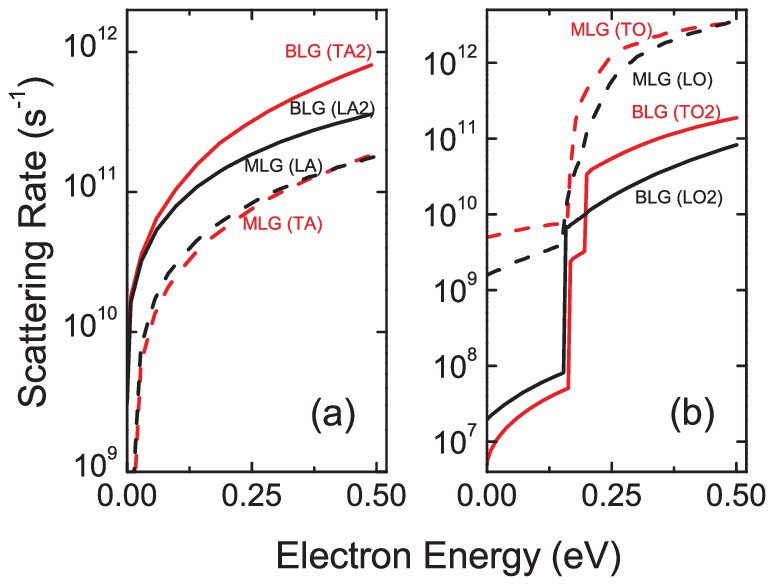}
\caption{(Color online) Intrinsic electron scattering rates in MLG and BLG calculated from the first principles for the dominant (a) acoustic and (b) optical phonons.~\cite{Borysenko2010, Borysenko2011} The sudden increases shown in (b) are due to the onset of optical phonon emission. They correspond to the phonon frequencies at the points of high symmetry in the first Brillouin zone: $\omega_{\Gamma} \approx$ 200 meV and $\omega_{K} \approx$ 160 meV.}
\label{ScaRate}
\end{figure}
\end{center}

The strength of the electron-phonon coupling is estimated from the
first principles by using the density functional theory.~\cite{Borysenko2010,Borysenko2011} A comparative analysis
reveals several qualitative differences in the intrinsic scattering
of MLG and BLG.  For one, MLG has six phonon branches with two carbon atoms in a unit cell, whereas these numbers double in BLG.  Then, BLG may need to consider both the intraband and interband transitions due to the presence of a close second conduction band $\pi_2^*$ in addition to the lowest $\pi_1^*$ states.  In addition, the optical phonons in BLG appear to be a relatively weak source of interaction unlike in MLG. Ultimately, the intrinsic scattering rate in BLG is dominated by the long wave acoustic phonons (intravalley scattering).  Figure~\ref{ScaRate} illustrates this general trend; only the dominant branches are shown for clarity of presentation. The nomenclature for the phonon modes can be found in
Refs.~\onlinecite{Borysenko2011} and \onlinecite{Borysenko2010}.

In the presence of a polar substrate, the graphene electron interaction with SPPs can play a significant role as the earlier studies in MLG have illustrated comprehensively.~\cite{Fratini2008,Konar2009,Li2010_2} A similar treatment can be extended to BLG.  By assuming that the electrons are equally distributed in the two layers of BLG, we can derive the corresponding scattering rate as
\begin{align}\label{SPPRate}
\frac{1}{\tau_{S}(\mathbf{k}_{i})} &=
\frac{2\pi}{\hbar}\sum_{\mathbf{q}}\frac{e^{2}\mathcal{F}^{2}}
{\varepsilon(q)^{2}}\left[\frac{e^{-2qd}+e^{-2q(d+c)}}{2q}\right] \nonumber \\
 &\times \left(n_q+
\frac{1}{2}\pm\frac{1}{2}\right)|g^{s,s^\prime}_{\mathbf{k}_i}(q)|^2 \delta(E_f-E_i\pm\hbar\omega_{S})
\,,
\end{align}
where $q=|\mathbf{k}_f-\mathbf{k}_i|$ is the magnitude of the SPP momentum,
$E_f$ ($E_i$) is the final (initial) electron energy, $d$ is the
distance between the first layer and the substrate (0.4 nm), $c$ is
the interlayer distance (0.34 nm), $\omega_{S}$ is the SPP energy,
$n_q$ is the phonon occupation number,
$\mathcal{F}^2=\frac{\hbar\omega_{S}}{2A\varepsilon_0}
(\frac{1}{\kappa_{S}^{\infty}+1}-\frac{1}{\kappa_{S}^{0}+1})$ is the
square of Fr\"{o}hlich coupling constant, and $\varepsilon(q)$ is
the dielectric function.  The term $|g^{s,s^\prime}_\mathbf{k}(q)|^2=\frac{1}{2}[1+ss^\prime\cos
{\alpha_\mathbf{k}}\cos{\alpha_\mathbf{k+q}}+ss^\prime\sin
{\alpha_\mathbf{k}}\sin{\alpha_\mathbf{k+q}}\cos{2\theta}]$ comes
from the overlap of the electron wave functions of the initial and
final states with the scattering angle $\theta$;~\cite{Wang2010} $s$ and $s^\prime$ are the band indices whose product is $+1$ for the intraband (for example, $\pi_1^*$-$\pi_1^*$) and $-1$ for the interband ($\pi_1^*$-$\pi_2^*$) transitions.~\cite{comment}  For intrinsic BLG, $\alpha_\mathbf{k}=\pi/2$ for an arbitrary $\mathbf{k}$. When an interlayer bias of $u$ is applied, it is modified to satisfy $\tan{\alpha_\mathbf{k}}=\hbar^2k^2/m^{\ast}u$, where $m^{\ast}$
is the effective mass of unbiased (or intrinsic) BLG at the $K$ or $K^\prime$ point.~\cite{Wang2010}  For MLG,
$|g^{s,s^\prime}_\mathbf{k}(q)|^2=\frac{1}{2}[1+ss^\prime\cos{\theta}]$.
The static dielectric function $\varepsilon(q)$ in BLG at room
temperature can be estimated by using the random phase approximation.
In addition, $\mathcal{F}$ contains the dependence on the high (low)
frequency dielectric constant of the substrate
$\kappa^{\infty}_{S}$ ($\kappa^{0}_{S}$). The specific values for the
relevant substrate parameters can be found in
Refs.~\onlinecite{Fratini2008}, \onlinecite{Konar2009}, and \onlinecite{Perebeinos2010}. As for the remote impurity scattering, the charged impurities are assumed to be located on the surface of the substrate, in which case the scattering rate is given by
\begin{align} \label{ImRate}
\frac{1}{\tau_{im}(\mathbf{k}_{i})} &=\frac{2\pi
n_i}{\hbar}\sum_{\bf{q}}\left[\frac{e^{2}}
{2\varepsilon_0\kappa\varepsilon(q)q}\right]^2\left[\frac{e^{-2qd} +e^{-2q(d+c)}}{2}\right]
\nonumber \\ &\times |g^{s,s^\prime}_\mathbf{k}(q)|^2\delta(E_f-E_i)
\,,
\end{align}
where $\kappa=(\kappa^{0}_{S}+1)/2$ is the background dielectric
constant and $n_i$ is the impurity concentration. The other
parameters are the same as defined in Eq.~(\ref{SPPRate}).

\begin{center}
\begin{figure}
\includegraphics[bb=243 514 416 672]{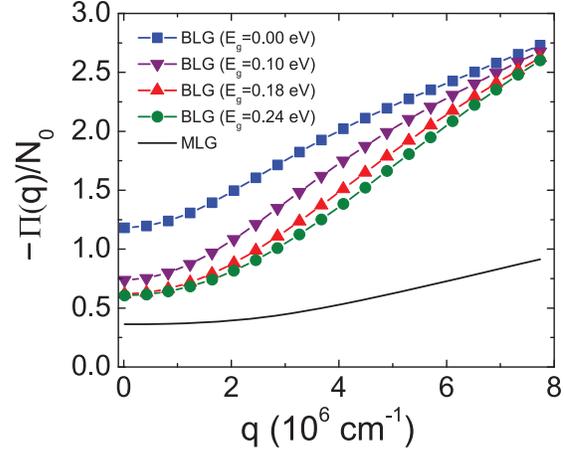}
\caption{(Color online) Electron-hole propagator $\Pi (q)$ normalized to $N_0$ ($ = 2m^*/\pi\hbar^2$) in MLG and BLG with the graphene electron density of $ 5 \times 10^{11}$~cm$^{-2}$.  For BLG, the calculation considers different interlayer biases with the induced energy gap $E_g$ of 0~eV (no bias), 0.10~eV, 0.18~eV, and 0.24~eV, respectively.} \label{propagator}
\end{figure}
\end{center}

The dielectric function in BLG can be expressed as~\cite{Sarma2008,Wang2010} $\varepsilon(q)=1-v_q\Pi(q)$ with the
bare Coulomb interaction $v_q=e^2/2\varepsilon_0q$ and the
electron-hole propagator
\begin{align} \label{Propagator}
\Pi(q) = 2 \sum_{s,s^\prime,\mathbf{k}}|g^{s,s^\prime}_\mathbf{k}(q)|^2
\frac{f_\mathbf{k+q}^{s^\prime}-f_\mathbf{k}^s} {E_\mathbf{k+q}^{s^\prime}-E_\mathbf{k}^s}
\,,
\end{align}
where $f_\mathbf{k}^s$ is the electron distribution function in band $s$ and  the factor of 2 takes into account the spin degeneracy. Figure~\ref{propagator} shows the numerically obtained propagators for MLG and BLG with the graphene electron density of $n=5\times10^{11}$~cm$^{-2}$ at 300 K. In this calculation [i.e., $\Pi(q)$], electron occupation in the $\pi_2^*$ states is ignored for its negligible contribution.  Compared to MLG, the screening in BLG appears to be much stronger due mainly to the large density of state at the bottom of the conduction band ($\pi_1^*$).~\cite{Sarma2008} Additionally, the impact of the interlayer potential on the electron screening in BLG can be substantial and is a strong function of temperature.

\section{Electron Transport in BLG Vs. MLG}

A full-band ensemble Monte Carlo calculation is used for evaluating
electron transport characteristics self-consistently. The model
takes into account the complete electron and phonon spectra in the
first Brillouin zone.  Specifically, both the graphene phonon
dispersion and its interaction with electrons are obtained from the
\textit{first principles} calculations as discussed
above,~\cite{Borysenko2010, Borysenko2011} whereas analytical
expressions from the tight binding approximation are used for the
electronic energy bands in MLG and BLG (with the nearest-neighbor
hopping energy $\gamma_0 = 3.3$~eV and and the interlayer hopping
energy $\gamma_1 = 0.4$~eV).~\cite{Li2009} The effect of degeneracy
in the electronic system is taken into account by the rejection
technique, after the final state selection.  Electron scattering
with the SPPs and remote impurities are also considered in the
calculation as described above whenever necessary.  Throughout the
calculation, we assume $n=5\times10^{11}$~cm$^{-2}$ and $T=300$~K.

\begin{center}
\begin{figure}
\includegraphics[bb=243 514 416 672]{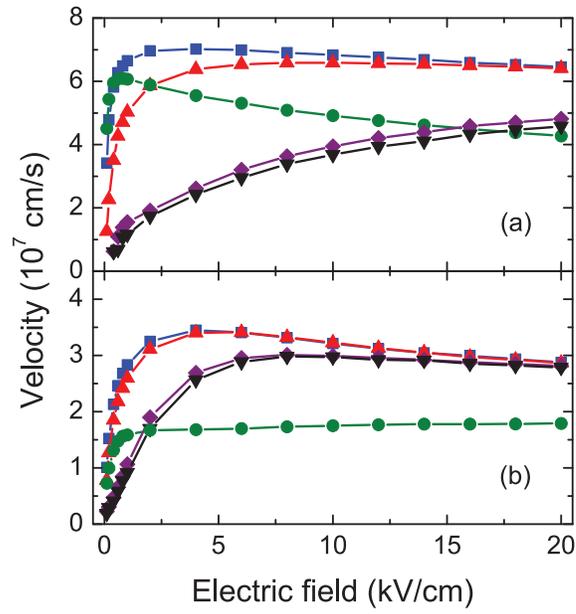}
\caption{(Color online) Electron drift velocity versus electric
field in (a) MLG and (b) BLG, with different substrate conditions: intrinsic/no substrate (circle),  SiO$_2$ (triangle), SiO$_2$ with impurities (reverse triangle), BN (square), and BN with impurities (diamond). The electron density is $5\times 10^{11}$~cm$^{-2}$ at 300~K. The impurities on the surface of the substrate (d=0.4 nm) have the density $5\times 10^{11}$~cm$^{-2}$.} \label{unbiased_VE}
\end{figure}
\end{center}

Figure~\ref{unbiased_VE} shows the electron drift velocity as a
function of the electric field in MLG and BLG (with no interlayer
bias).  It provides a comparison of all examined cases: namely,
intrinsic graphene and graphene on two different substrates, SiO$_2$
and BN, for which a charged impurity density of
$n_i=5\times10^{11}$~cm$^{-2}$ is considered along with the SPPs. As
illustrated, intrinsic MLG (i.e., no substrate) shows remarkable
transport properties,~\cite{Li2010,Li2010_2} with the mobility and
the saturation velocity as high as $1.0\times10^{6}$~cm$^{2}$/Vs and
$4.3\times10^{7}$~cm/s. An analogous calculation for BLG gives a
substantially lower mobility of $1.2\times10^{5}$~cm$^{2}$/Vs and
the saturation velocity of $1.8\times10^{7}$~cm/s.  These results
for the intrinsic drift velocity are consistent with the scattering
rates in Fig.~\ref{ScaRate}; they demonstrate how acoustic and
optical phonons affect the charge carriers in MLG and BLG at various
electric fields.



\begin{center}
\begin{figure}
\includegraphics[bb=243 514 416 672]{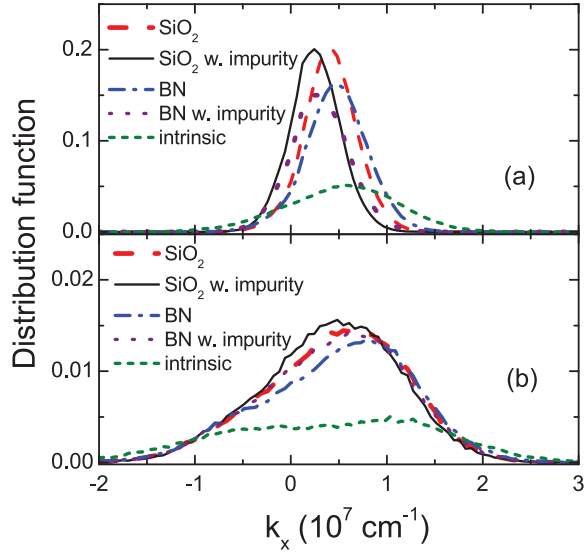}
\caption{(Color online) Cross sectional view ($k_y$=0) of electron distribution functions in (a) MLG and (b) BLG at 20~kV/cm, with different substrate conditions: intrinsic/no substrate (short dashed line), SiO$_2$ (long dashed line), SiO$_2$ with impurities (solid line),  BN (dashed-dotted line), and BN with impurities (dotted line).  The conditions are the same as specified in Fig.~\ref{unbiased_VE}.} \label{distribution}
\end{figure}
\end{center}

The lower mobility in BLG can be readily explained by the larger
acoustic phonon scattering rates, as well as the lower electron
speed near the bottom of the conduction band where all electrons
tend to congregate at low electric fields. To understand the smaller
saturation velocity in BLG, on the other hand, it is convenient to examine the distribution function at high electric field plotted in
Fig.~\ref{distribution}.  As the electrons gain energy in the
applied field, the distribution function shifts in the k-space along the direction of the drift. At the same time, increased scattering with the long-wave acoustic phonons leads to a further broadening of the electron distribution. The result of this interplay between the displacement
and the broadening is the saturation of the drift velocity in intrinsic BLG, where the acoustic phonons dominate the scattering.  In MLG, however, the velocity curve appears to demonstrate another pattern, which points to the
presence of a competing intrinsic scatterer. Indeed, as we discussed
earlier, the inelastic scattering via optical phonons is strong unlike in BLG, providing energy relaxation for hot electrons.  Consequently, the distribution in MLG is prevented from shifting further towards higher energies, which in turn effectively curtails the momentum relaxing interactions and results in a higher saturation velocity.

A similar consideration applies when an additional source of optical phonon scattering comes into play.  That is, the saturation velocity may be actually enhanced when graphene electrons are subject to the SPP scattering, as it provides another path for hot electron energy relaxation.
Figure~\ref{unbiased_VE} clearly demonstrates this effect. In MLG, the SPP scattering can increase the saturation velocity up to
$6.5\times10^{7}$~cm/s if SiO$_2$ or BN is used for the substrate
material. This trend of the positive impact of the substrate on the
saturation velocity was confirmed in recent
studies.~\cite{Perebeinos2010,Ashley2010,Li2010_2} In BLG, the
saturation velocity can reach $2.9\times10^{7}$~cm/s, which is about
as 1.5 times large as the intrinsic value. Even though the screening
in BLG is larger, leading to smaller SPP scattering rates, the
enhancement of drift velocity is still prominent.  This is due partly to the fact that the competing relaxation mechanism (i.e., intrinsic optical phonon scattering) is weak in BLG as discussed earlier.


When the charged impurity scattering is taken in account, the
electron drift velocities in MLG and BLG are substantially reduced
at low electric fields (see Fig.~\ref{unbiased_VE}). If
BN is used as the substrate, the low field mobility is
$1.9\times10^{4}$~cm$^{2}$/Vs in MLG, and
$1.2\times10^{4}$~cm$^{2}$/Vs in BLG. The drift velocity at the
electric field of 20 kV/cm in MLG decreases to
$4.8\times10^{7}$~cm/s, while in BLG it stays nearly the same as the
case without impurity scattering, $2.8\times10^{7}$~cm/s. Clearly,
the impact of ionized impurity scattering is much less severe in BLG.  This is because the BLG electrons on average are more separated from the impurities (i.e., the surface of the substrate) and experience stronger screening.  The calculated result in MLG on SiO$_2$ shows a good agreement with the available experimental data;~\cite{Ashley2010} the discrepancy may be attributed to the presence of additional sources of interaction such as neutral scatters on the substrate, ripples and other defects in graphene crystal lattice.~\cite{Neto2009}

\section{Transport in BLG with interlayer bias}
\begin{center}
\begin{figure}
\includegraphics[bb=243 514 416 672]{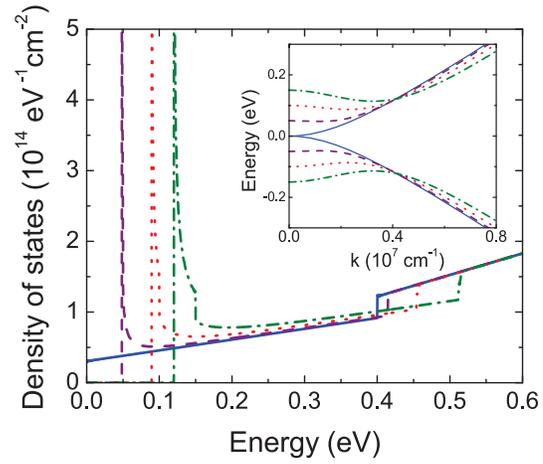}
\caption{(Color online) Density of states in BLG for different
interlayer biases with the induced energy gap $E_g$ of 0~eV (solid
line), 0.10~eV (dashed line), 0.18~eV (dotted line), and 0.24~eV
(dashed-dotted line), respectively. In the insert, the corresponding
band structures are plotted. } \label{dos}
\end{figure}
\end{center}

\begin{center}
\begin{figure}
\includegraphics[bb=243 514 416 672]{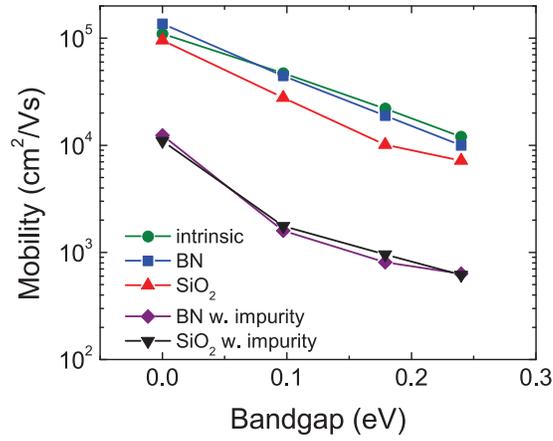}
\caption{(Color online) Electron mobility versus bias-induced
bandgap in BLG at 300~K, with different substrate conditions. The
electron density is $5\times 10^{11}$~cm$^{-2}$. The impurities on
the surface of the substrate (d=0.4 nm) have the density $5\times
10^{11}$~cm$^{-2}$. }\label{mobility}

\end{figure}
\end{center}

BLG with an interlayer bias offers the advantage of a tunable bandgap with potential applications to transistors, tunable photo-detectors
and lasers.~\cite{Castro2007,Xia2010}  The band structure in this case changes to,~\cite{McCann2006}
\begin{align}\label{bandstrucuture}
E_k^2=\frac{\gamma_1^2}{2}+\frac{u^2}{4}+\hbar^2v_F^2k^2\pm
\sqrt{\frac{\gamma_1^4}{4}+\hbar^2v_F^2k^2(\gamma_1^2+u^2)}\,,
\end{align}
where $u$ is the difference between on-site energies in the two
layers, $v_F=(\sqrt{3}/2)a\gamma_0/\hbar\approx1\times10^8$ cm/s is
the Fermi velocity ($a=0.246$~nm), and the minus and plus signs
correspond the $\pi_1^*$ and $ \pi_2^*$ conduction bands,
respectively.  However, this ability to tune the band gap comes at
the expense of the material's intrinsic transport properties. As the
bandgap opens, the bottom of the lowest conduction band changes its
shape from a hyperbola to a so-called "Mexican hat" and the density
of states exhibits a Van Hove singularity as shown in
Fig.~\ref{dos}. The inserted figure shows the band structures for
different bandgaps. Consequently, the increased density of states
leads to stronger quasi-elastic electron interaction with the
long-wavelength acoustic phonons at low electron energies.
Figure~\ref{mobility} provides the dependence of the mobility on the
size of the bandgap. When the bandgap is $0.24$ eV, the mobility
drops as low as to $1.2\times10^4$~cm$^{2}$/Vs even without any
external scattering mechanisms, which is one order of magnitude
smaller than in the unbiased case.  The reduction in the mobility
becomes even more pronounced when the ionized impurity scattering is
taken into account.  The Van Hove singularity also enhances the
impact of electron coupling with the remote impurities.

\begin{center}
\begin{figure}
\includegraphics[bb=243 450 416 672]{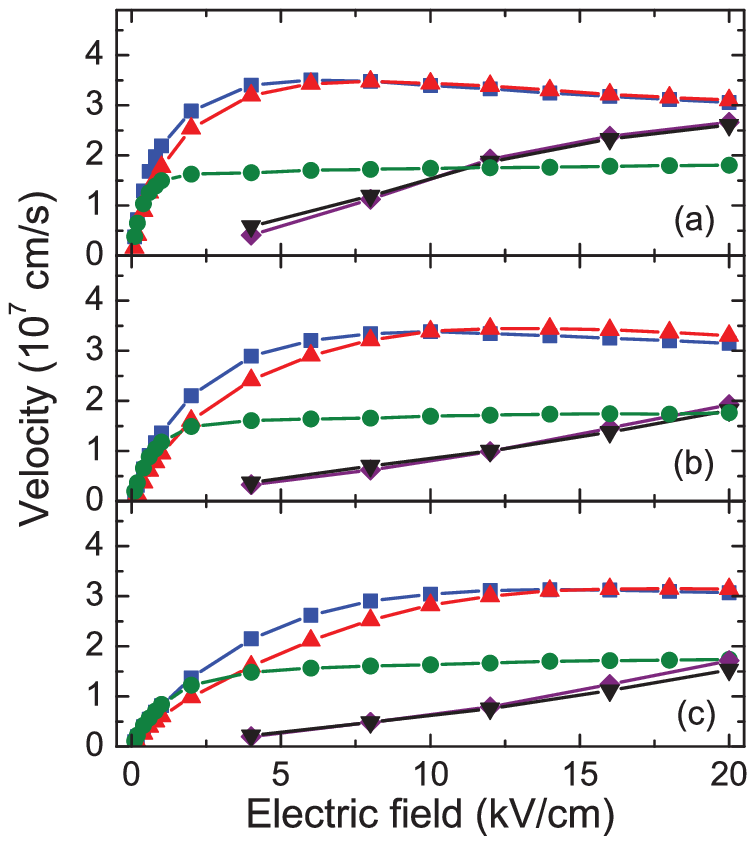}
\caption{(Color online) Electron drift velocity versus electric field in BLG with a bias-induced gap of $E_g$ of (a) 0.1~eV, (b) 0.18~eV, and (c) 0.24~eV, under different substrate conditions: intrinsic/no substrate (circle), SiO$_2$ (triangle), SiO$_2$ with impurities (reverse triangle), BN (square), and BN with impurities (diamond). The conditions are the same as specified in Fig.~\ref{mobility}.} \label{biased_VE}
\end{figure}
\end{center}

On the other hand, it appears that the velocities at high fields are mostly unaffected by the gap.  Figure~\ref{biased_VE} shows the
drift velocity for $u= 0.1$~V, $0.2$~V, and $0.3$~V, which correspond to the bandgap of $0.1$~eV, $0.18$~eV, and $0.24$~eV, respectively. As the velocity saturation is associated with hot electrons, the changes of the electron spectrum on the bottom of the conduction band have little to no effect on the charge carriers located much higher in energies. Similarly, the impact of the SPP scattering on the drift velocity, as it is felt primarily via the high energy electrons, is not affected by the gap in the electron energy spectrum. Figure~\ref{biased_VE} demonstrates this point as no appreciable difference is observed in the SPP contribution for three different choices of $u$.  When the ionized impurity scattering is included, the velocity saturation is progressively pushed to a higher field due to the reduction in the mobility (i.e., the slope) that was discussed above.

It should be noted that the mobilities obtained in the presented
Monte Carlo simulation may have a limited accuracy, for lack of an
available adequate model for the electron-electron scattering. It
might be especially important in case of biased BLG, where the
divergence of the density of states results in difficulties of
reproducing electron thermal equilibrium in Monte Carlo simulation.

\section{Summary}
Electron transport properties of BLG are studied under realistic conditions in the presence of the SPPs and charged impurities. Overall, BLG has a lower mobility and saturation velocity than MLG, due to the stronger acoustic phonon
scattering, weaker optical phonon scattering, and nonlinear dispersion at the bottom of the conduction band. It is also shown that SPPs can improve the saturation velocity in BLG by effectively dissipating the electron energy. The impurity scattering has a strong effect of decreasing the drift velocities in both MLG and BLG. However, BLG appears more resistant to impurity scattering
than MLG, due to a stronger screening and larger effective distance
between electrons and the impurities. In BLG with a interlayer bias, the changes in the band structure result in drastically reduced mobilities,
especially in the presence of charged impurities.  This may limit the chances of utilizing BLG in device applications.

\begin{acknowledgments}
This work was supported, in part, by the DARPA/HRL CERA, ARO, and
SRC/FCRP FENA programs. MBN wishes to acknowledge partial support
from the Office of Basic Energy Sciences, US DOE at Oak Ridge
National Lab under contract DE-AC05-00OR22725 with UT-Battelle, LLC.
\end{acknowledgments}

\clearpage

\clearpage

\clearpage

\clearpage

\clearpage

\clearpage

\clearpage



\begin{references}
\bibitem{comment} This expression for $g^{s,s}_{\mathbf{k}}(q)$ is also applied to the $\pi_2^*$-$\pi_2^*$ scattering even though it is only approximate. the treatment does not alter the calculation results due to the small contribution of the electrons in the $\pi_2^*$ band.
\end{references}
\end{document}